\documentclass{article}

\begin{document}

\title{Lagrangian Approach to Quantum Mechanics}

\author{Yong Gwan Yi}

\maketitle

\begin{abstract}
The Lagrangian approach of Dirac is presented in a complete 
form. This suggests to identify the Schr\"{o}dinger equation 
as the Euler-Lagrange equation rather than the Hamiltonian 
operator equation.
\end{abstract}

\bigskip
\bigskip

Planck found the relation of frequency to energy and de Broglie 
the relation of wavelength to momentum. In the wave-particle 
duality, the phase of matter waves could be written as
\begin{equation}
(i/\hbar)(px-Et)\quad\mbox{from}\quad
i(kx-\omega t).
\end{equation}
Schr\"{o}dinger deduced the wave equation for a particle from 
the Hamiltonian of a classical system. The Schr\"{o}dinger 
equation led to the formulation of quantum mechanics and the
Hamiltonian came to be the argument of correspondence [1].

Dirac took up the question of what corresponds in the quantum 
theory to the Lagrangian method of classical theory [2]. 
Dirac tried to take over the idea rather than the equation, 
provided by the Lagrangian. But the action principle involves 
the Euler-Lagrange equation, which is identical with the 
Schr\"{o}dinger equation. In this paper, I have tried to 
complete the Lagrangian approach of Dirac in relation to the 
Schr\"{o}dinger equation. For the continuity of discussion, 
arguments are restricted to one-dimensional cases.

The Lagrangian approach begins by writing the phase of matter 
waves in the integral form
\begin{equation}
(i/\hbar)(pdx-Edt),\quad\mbox{so}\quad
(i/\hbar)L(x,\dot{x})dt.
\end{equation}
In the quantum theory of canonical coordinate and momentum, 
however, we cannot take over the classical notion of coordinate 
and velocity. Therefore, we must consider the Lagrangian as a 
function of $x$ at time $t$ and $x+dx$ at time $t+dt$ rather 
than $x$ and $\dot{x}$. Dirac has used the action function $S$ 
for the time integral of the Lagrangian and the action principle. 
But the integral is called Hamilton's principal function and 
Hamilton's principle in classical mechanics [3].

In the theory of representations, the transformation 
function connects two observables for the state vectors [4]. 
The notation $<x|x'>$ is given for the function when 
observations are made of coordinates. Dirac showed that
\begin{equation}
<x|x'>\quad\mbox{corresponds to}\quad 
\exp\{iS(x,x')/\hbar\}.
\end{equation}
We may divide up the interval into a number of small 
sections by the introduction of a sequence of 
intermediate times $t_i$. Then we have for $S$
\begin{equation}
S(x_n,x_1)=S(x_n,x_{n-1})+\cdots+S(x_2,x_1).
\end{equation}
The corresponding function becomes
\begin{equation}
<x_n|x_1>=\int\cdots\int<x_n|x_{n-1}>dx_{n-1}
\,\cdots\,<x_3|x_2>dx_2<x_2|x_1>.
\end{equation}
We must integrate over the intermediate $x_i$ between 
transformations, which follows from the property of state 
vectors. The composition law (5) can equally be expressed 
in the form of a recursive relation
\begin{equation}
<x_{i+1}|=\int<x_{i+1}|x_i>dx_i<x_i|
\quad\mbox{for}\quad i=2,3,\cdots,n-1.
\end{equation}
They describe the development of a wave function with time 
in the general form
\begin{equation}
<x|=\int<x|x'>dx'<x'|\quad\mbox{or}\quad
\psi(x)=\int<x|x'>dx'\,\psi(x').
\end{equation}
In the form of integral equation, $<x|x'>$ is the development 
of a wave function given at $x'$ at an earlier time $t'$ 
into a wave function at any other point $x$ at a time $t$. 
Dirac explained, $<x|x'>$ is that solution of the 
Schr\"{o}dinger equation. Feynman remarked, it is a kind 
of Green’s function for the Schr\"{o}dinger equation. 
Feynman developed the path integral formulation based on  
equation (7) [5].

In the classical theory, the motion of a system from time 
$t'$ to time $t$ is such that the action function is stationary 
for the path of motion. The action principle is a statement 
that the variation of $S$ from time $t'$ to time $t$ is zero: 
$\delta S=0$. The classical requirement that the values of 
the intermediate $x_i$ shall make $S$ stationary corresponds 
to the condition in the quantum theory that all values of 
the intermediate $x_i$ are important in proportion to their 
contribution to the integral. This shows the way in which 
the action principle is absorbed in the composition law. 
We can summarize Dirac's remark on the action principle 
by requiring the variation of $<x|x'>$ to be zero in (3):
\begin{equation}
\delta<x|x'>=0 \quad\mbox{corresponds to}\quad
\delta S=0 \quad\mbox{from}\quad
\delta e^{iS/\hbar}=(i\delta S/\hbar)e^{iS/\hbar}. 
\end{equation}
The action principle is a statement about the integral of 
$L$, from which we obtain the solution of the problem via 
the Euler-Lagrange equation. Here, the variational problem 
is to determine $<x|x'>$ that will make the development of 
a wave function stationary. When viewed from the present 
point, the Schr\"{o}dinger equation is the condition for 
a wave function to take on a stationary distribution.

In the classical theory, $S$ is the generating 
function of a contact transformation relating $(x,p)$ 
to $(x',p')$:
\begin{equation}
\frac{\partial}{\partial x}S(x,x')=p,\quad
-\frac{\partial}{\partial x'}S(x,x')=p'.
\end{equation} 
In the quantum theory, it becomes the generator of a 
unitary transformation connecting the two state 
vectors: 
\begin{equation}
\frac{\hbar}{i}\frac{\partial}{\partial x}
<x|x'>=p<x|x'>,\quad
-\frac{\hbar}{i}\frac{\partial}{\partial x'}
<x|x'>=p'<x|x'>.
\end{equation}
The quantum equation for $<x|$ or $|x'>$ is given by 
multiplying with $<x'|$ or $|x>$ and integrate over $x'$ 
or $x$. Written in terms of wave function, they are
\begin{equation}
\frac{\hbar}{i}\frac{\partial\psi}{\partial x}=p\psi,\quad
-\frac{\hbar}{i}\frac{\partial\psi^*}{\partial x}=p\psi^*.
\end{equation}

We can extend the analogue of transformation to the equation 
of motion. This we do by applying the idea of Dirac to the 
Hamilton-Jacobi equation. The relation to the Schr\"{o}dinger 
equation is the center of attention. The Hamilton-Jacobi 
equation is 
\begin{equation}
\frac{\partial S}{\partial t}+H=0.
\end{equation}
Thus, the quantum analogue has the form
\begin{equation}
\frac{\hbar}{i}\frac{\partial}{\partial t}
<x|x'>+H<x|x'>=0.
\end{equation}
From the form of the Hamiltonian, the Hamilton-Jacobi 
equation can be written
\begin{equation}
\frac{\partial S}{\partial t}+
\frac{1}{2m}\biggl(\frac{\partial S}{\partial x}
\biggr)^2+V=0.
\end{equation}
Then, the quantum analogue has the form
\begin{equation}
\frac{\hbar}{i}\frac{\partial}{\partial t}
<x|x'>+\frac{1}{2m}\biggl(\frac{\partial S}{\partial 
x}\biggr)^2<x|x'>+V<x|x'>=0.
\end{equation}
The quantum equation for $<x|$ follows from the form of (15), 
as can be seen when we multiply the equation by $<x'|$ and 
integrate with respect to $x'$. The quantum equation is 
essentially the same as used for the derivation of the 
Schr\"{o}dinger equation. In putting into the Schr\"{o}dinger 
equation, however, the momentum should be treated as a 
constant. In general,
\begin{equation}
\biggl(\frac{\hbar}{i}\frac{\partial}{\partial x}
\biggr)^2<x|x'>=\frac{\hbar}{i}\frac{\partial}
{\partial x}\biggl(\frac{\partial S}{\partial x}
<x|x'>\biggr)=\frac{\hbar}{i}\frac{\partial^2S}
{\partial x^2}<x|x'>+\biggl(\frac{\partial S}
{\partial x}\biggr)^2<x|x'>.
\end{equation}
In deriving the Schr\"{o}dinger equation, actually, the 
momentum has been treated as a constant. 

The potential scattering is represented by a potential 
energy which is appreciably different from zero only 
within a finite region. The Hamilton-Jacobi equation 
for scattering can then be written as
\begin{equation}
\frac{1}{2m}\biggl(\frac{\partial S}{\partial x}
\biggr)^2-H_0=\left\{\begin{array}{ll}
-V(x)&\mbox{if }x\rightarrow x'\\
0&\mbox{otherwise}\end{array}\right.,
\end{equation}
where $H_0$ is the energy of the free particle. From 
this form of the Hamilton-Jacobi equation we may 
deduce the corresponding quantum equation of the form
\begin{equation}
\frac{1}{2m}\biggl(\frac{\partial S}{\partial x}
\biggr)^2<x|x'>-H_0<x|x'>=-V(x)\delta(x-x').
\end{equation}
The inhomogeneity of potential energy is the reason 
for the choice of the $\delta$ function instead of the 
transformation function. As the momentum of the free particle 
is a constant, we can put the quantum equation into the form
\begin{equation}
-\frac{\hbar^2}{2m}\frac{\partial^2}{\partial x^2}
<x|x'>-H_0<x|x'>=-V(x)\delta(x-x').
\end{equation} 
The Hamilton-Jacobi equation for scattering gives the integral 
form of the Schr\"{o}dinger equation. This is identical with 
the integral equation which was introduced by Born from the 
Schr\"{o}dinger equation for scattering. Here, the quantity 
$<x|x'>$ has been a link relating the two equations, 
differential and integral.

We are taught the correspondence between the Schr\"{o}dinger 
equation and the Hamiltonian of classical systems. But their 
correspondence is not complete. Because the differential 
operator acts on everything that stands to the right, the 
same relation does not always hold between operators as between 
variables. As can be seen in (16), the second application of 
the momentum operator to a wave function gives rise to the 
differentiation of $p$ in addition to the general effect of 
multiplication of the wave function by $p^2$. If the momentum 
is a constant, there is no additional term. This is the case 
for the hydrogen atom problem where the angular momentum is 
the quantum number. But it is actually so in the harmonic 
oscillator problem, for the momentum depends linearly on its 
coordinate. In fact, the zero-point energy is a result of the 
additional term in the Schr\"{o}dinger equation of harmonic 
oscillator. It characterizes a difference of correspondence 
with the Hamiltonian of classical system. The zero-point 
energy was also noted by Heisenberg in his quantum theoretical 
calculations [6]. The Schr\"{o}dinger equation needs to be 
justified in a different way. 

Their correspondence is not complete either in form. It is 
the expectation value in the quantum theory that corresponds 
to the Hamiltonian of classical theory. As shown by Ehrenfest 
[7], the equation for physical quantities can be supposed to 
be the equation for the expectation values in the quantum 
theory. The Hamiltonian required by the transition from 
classical to quantum physics should therefore have the form
\begin{equation}
\int<x|\frac{\hbar}{i}\frac{\partial}{\partial t}
|x>dx+\int<x|\frac{1}{2m}\biggl(\frac{\partial S}
{\partial x}\biggr)^2|x>dx+\int<x|V|x>dx=0.
\end{equation}
This follows for an arbitrary $|x'>$ from the very form 
of (15) when we multiply the equation by $|x>$ and integrate 
over $x$. Written in terms of wave function, the quantum 
theoretical Hamiltonian reads 
\begin{equation}
\int\psi^*E\psi\,dx=
\int\biggl\{\frac{1}{2m}
\biggl(-\frac{\hbar}{i}\frac{\partial\psi^*}
{\partial x}\biggr)\biggl(\frac{\hbar}{i}
\frac{\partial\psi}{\partial x}\biggr)
+\psi^*V\psi\biggr\}\,dx.
\end{equation}
This is known as an example of variational problem provided 
by the Schr\"{o}dinger equation [8]. The variational approach 
is more than just a matter of academic curiosity. The form of 
expression is just what we should take in the quantum theory  
for the Hamiltonian of classical system. By the action 
principle, actually, it has shown how the Hamiltonian of 
quantum system goes over into the Schr\"{o}dinger equation. 
Here, the action principle becomes a statement that the energy 
of quantum system is a constant: $\delta<E>=0$. 

According to Dirac, the action principle comes to be the 
argument of correspondence of the classical to quantum 
equations of motion. In fact, it is the variational 
expression in (21) that corresponds completely to the 
Hamiltonian of classical system. In the Schr\"{o}dinger 
equation has the differentiation of $p$ been involved 
unintentionally. If the momentum is not a constant, the 
additional term is unavoidable. It does not matter if the 
term is included in the equation of motion. But it does 
in the Hamiltonian. This is a problem of justification. 
In form and content, it is reasonable to identify the 
Schr\"{o}dinger equation as the Euler-Lagrange equation 
obtained from the variational principle for the energy of 
quantum system rather than the Hamiltonian operator equation. 

\bigskip

\centerline{\bf Appendix: The zero-point energy in 
calculation}

\bigskip

The linear harmonic oscillator is the one-dimensional 
motion of a point mass $m$ attracted to an 
equilibrium position $x=0$ by a force that is 
proportional to the displacement $x$ from it. The 
restoring force can be represented by the potential 
energy $V(x)=m\omega^2x^2/2$. Introducing the 
variable $\xi=(m\omega/\hbar)^{1/2}x$ and the 
eigenvalue $\lambda=2E/\hbar\omega$, we can put the 
Schr\"{o}dinger equation in the form
\begin{equation}
\frac{d^2\psi}{d\xi^2}+(\lambda-\xi^2)\psi=0.
\end{equation}
For sufficiently large $\xi$, the dominant behavior
of $\psi$ is given by $\exp(-\xi^2/2)$. Thus, we can 
assume an exact solution of the form $H(\xi)\exp(-\xi^2/2)$. 
In the process of solving (21), we have used for the 
asymptotic part the relations 
\begin{equation}
\frac{d\psi}{d\xi}=-\xi\psi\quad\mbox{and}\quad
\frac{d^2\psi}{d\xi^2}=\xi^2\psi-\psi.
\end{equation}
The first shows a typical operator-eigenvalue relation. 
But the second gives an additional term, a result of 
differentiating $\xi$. This corresponds formally to the 
differentiation of momentum itself, resulting in the 
zero-point energy.

\end{document}